\documentclass[aps,prl,twocolumn,groupedaddress,showpacs]{revtex4}
\usepackage{epsfig}
\usepackage{amsmath}
\usepackage{amsfonts}
\def\bE{\mathbf E}

\def\bH{\mathbf H}
\def\bJ{\mathbf J}
\def\bB{\mathbf B}
\def\bD{\mathbf D}

\def\bx{\mathbf x}

\def\bw{\mathbf w}

\def\b0{\mathbf 0}

\begin{document}

\title{Exponential cosmological redshift in a linearly expanding universe}
\author{Neil V. Budko}
\affiliation{Laboratory of Electromagnetic Research, Faculty of Electrical
Engineering, Mathematics and Computer Science,
Delft University of Technology,
Mekelweg 4, 2628 CD Delft, The Netherlands}
\email{n.v.budko@tudelft.nl}

\date{\today}

\begin{abstract}
The first principles analysis of the radiation by an arbitrary source in a flat Friedmann-Robertson-Walker space-time is presented.
The obtained analytical solution explicitly shows that the cosmological redshift is not of kinematic origin and that the source and the observer 
may be regarded as being at rest with respect to eachother at all times.
At the same time the effect of the time-variation of the metric on the propagation of light appears to be underestimated
in the standard cosmology. The cosmological redshift caused by the linear time-variation of the metric turns out to be an exponential 
rather than linear function of the well-defined spatial distance and the apparent brightness of the source contains an even 
stronger exponential decay factor.
\end{abstract}

%\keywords{}

\maketitle

The distance-dependent redshift of the electromagnetic radiation from stellar sources is the key observational 
parameter in astronomy and a reality-check for any viable cosmological theory. The expanding universe
model described in general relativity by the Friedmann-Robertson-Walker (FRW) metric \cite{CosmologyBook} 
is accepted as the most 
natural explanation of the observed data, i.e., the Hubble's approximately linear law. This law has been deduced experimentally  
by estimating the distances to particular types of redshifted stellar sources -- standard astronomical candles -- 
from the measurement of their apparent brightness \cite{HubblePaper}. The preferred theoretical explanation goes 
back to Lemaitre \cite{Lemaitre} and is based on the analysis of the light-cone condition for the space-time interval 
in the FRW metric. However, the original {\it linear} Hubble's redshift law is recovered only in the ``small'' 
distance (small redshift) approximation \cite{LandauLifshitz}. An exact law universaly valid for large distances 
and large redshifts has never been derived. Recently we have seen a surge in the astronomical data containing 
extremely high redshifts \cite{HighRedshift}. There is also a growing interest in various new cosmological scenarious, 
e.g. inflation, accelerated expansion, varying speed of light theories, and cyclic universes \cite{NewCosmologies}.
Since all of these theories and data need to be tested against the simplest flat-space linearly expanding universe model,
the exact redshift and apparent brightness (or luminocity) formulas resulting from this basic cosmological 
model and valid for all distances have become especially important.

On the other hand, the physical meaning of the expanding universe model is also still open to debate \cite{Debate}.
What exactly is expanding? Does the spatial separation between galaxies actually grow in time?
How can this mutual recession be superluminal (for higher redshifts)? These elementary questions
seem to have as many answers as there are cosmologists, with the majority favouring
a more or less literal picture of galaxies flying apart. This is often appended by a vague statement 
that they are not flying apart {\it in} space, 
but rather because the space itself is expanding. 
This interpretation is an important part of the {\it standard model of
cosmology}. In particular, the literal expansion interpretation means that in the past all objects were 
physically closer to eachother, and that there was a very densely packed state in the 
beginning -- the Big Bang. 
 
Being a very simple algebraic expression, the space-time interval, which forms the backbone 
of the modern cosmological analysis, unfortunately, allows conflicting interpretations still causing many 
confusions and heated debates \cite{Debate}. 
Even the most basic conclusions are not as clearcut as it seems. For example, being shown the interval in the 
flat expanding FRW metric 
 \begin{align}
 \label{eq:Interval}
 {\rm d}s^{2}=c^{2}{\rm d}t^{2}
 -a^{2}(t)\left({\rm d}x_{1}^{2}+{\rm d}x_{2}^{2}+{\rm d}x_{3}^{2}\right),
 \end{align}
we usually hear something like this: The change of the scale factor $a(t)$ with time, e.g. linear growth, means
that all spatial distances are increasing with time. Hence, all objects will be receeding from 
eachother. This seemingly obvious statement, in fact, does not follow from (\ref{eq:Interval}) at all.
What we see in (\ref{eq:Interval}) is that the spatial interval is changing with respect to the
{\it time-interval}. Whether this means an increase of the spatial distance
with {\it time} as well remains to be seen. In particular, does the {\it co-moving} distance $a(t)R$ have any
physical meaning?

Here and in the companion paper \cite{Budko2009b} devoted to the wavepropagation in non-autonomous time-varying media 
I describe a general first-principles approach which may provide a definitive resolution of this and other interpretational problems. 
I obtain the exact redshift-distance relation valid for all distances and show the way to derive the 
exact brightness (luminocity) formula. The basic idea goes back to Schr{\"o}dinger and amounts to first 
posing and solving a general-relativistic problem for an arbitrary electromagnetic source radiating into a 
homogeneous infinite background with an FRW metric, and only then trying to find the physical interpretation. 
Fortunately, the solution is analytical and its complete derivation can be found in \cite{Budko2009b}. 
However, it does have several surprising features, which in my view call for a second look at our redshift data 
and may even require a revision of the standard intepretation mentioned above. 
First of all, the solution explicitly shows that the cosmological redshift is not of kinematic origin.
Secondly, the influence of a time-varying metric on the electromagnetic field seems to be
severely underestimated in the standard cosmology. Not only the exact Hubble law turns out to be 
exponential rather than linear, but there is also a very strong exponential decay in the
apparent brightness of the source. The latter is heavily relied upon in the verification of the Hubble law and to 
find possible deviations from its linear form for higher redshifts.

The earliest and, probably, the most complete attempt along the present lines 
is the Schr{\"o}dinger's 1939 paper \cite{Schrodinger1939}. 
Schr{\"o}dinger, however, considered the covariant {\it scalar} wave equation without the source term,
which, strictly speaking, is not the equation governing the electromagnetic radiation by a causal source.
The first-principles analysis should start with the covariant first-order Maxwell's system, and
proceed by {\it deriving} the corresponding wave equation, including the source-term, 
and solving it. Although, the first-order formulation of the probem has been considered many times since
\cite{Plebanski1960, deFelice1971, Bahram1973}, no deviations from the original linear 
Hubble's law were reported. 

In Cartesian coordinates the FRW metric of a spatially flat expanding universe is 
\begin{align}
\label{eq:Metric}
\left[g_{\alpha\beta}\right]=
\begin{bmatrix}
1 & 0 & 0 & 0\\
0 & -a^{2}(t) & 0 & 0\\
0 & 0 & -a^{2}(t) & 0\\
0 & 0 & 0 & -a^{2}(t)
\end{bmatrix},
\end{align}
with determinant $g=-a^{6}(t)$, where $a(t)$ is the cuvature radius determining the expansion rate.
As was demonstrated by Tamm \cite{Tamm1924,Plebanski1960,Leonhardt2006}, the electromagnetic field in an empty, but 
curved space-time, at each location $\bx$ and with respect to the
local time $t$ satisfies the usual Maxwell's equations in the $(3+1)$-form
\begin{align}
\label{eq:Maxwell}
\begin{split}
-\nabla\times\bH(\bx,t)+\partial_{t}\bD(\bx,t)&=-\sqrt{-g}\bJ(\bx,t),
\\
\nabla\times\bE(\bx,t)+\partial_{t}\bB(\bx,t)&=0,
\end{split}
\end{align}
with the following metric-induced local constitutive relations:
\begin{align}
\label{eq:ConstitutiveG}
\begin{split}
\bD(\bx,t)&=\varepsilon_{0}\varepsilon\bE(\bx,t)+\frac{1}{c}\bw\times\bH(\bx,t),
\\
\bB(\bx,t)&=\mu_{0}\mu\bH(\bx,t)-\frac{1}{c}\bw\times\bE(\bx,t),
\end{split}
\end{align}
where 
\begin{align}
\label{eq:Epsilon}
\varepsilon=\mu=-\frac{\sqrt{-g}}{g_{00}}g^{ij},
\;\;\;
w_{i}=\frac{g_{0i}}{g_{00}},
\end{align}
with $i,j=1,2,3$, and $\left[g^{\alpha\mu}\right]=\left[g_{\alpha\mu}\right]^{-1}$.
For the FRW metric (\ref{eq:Metric}) these relations reduce to
\begin{align}
\label{eq:Constitutive}
\begin{split}
\bD(\bx,t)&=\varepsilon_{0}a(t)\bE(\bx,t),
\\
\bB(\bx,t)&=\mu_{0}a(t)\bH(\bx,t),
\end{split}
\end{align}
Substituting these into (\ref{eq:Maxwell}) we arrive at the following Maxwell's equations:
\begin{align}
\label{eq:MaxwellReduced}
\begin{split}
-\nabla\times\bB(\bx,t)+\mu_{0}a(t)\partial_{t}\bD(\bx,t)&=-\mu_{0}a^{4}(t)\bJ(\bx,t),
\\
\nabla\times\bD(\bx,t)+\varepsilon_{0}a(t)\partial_{t}\bB(\bx,t)&=0.
\end{split}
\end{align}
Let us clarify the physical meaning of our coordinates. The Maxwell equations 
(\ref{eq:MaxwellReduced}) describe the {\it local} behaviour of the electromagnetic field.
The time-variable involved is called the {\it synchronous time} \cite{LandauLifshitz}, meaning 
that any physical process, say, emission of $N$ photons, which takes $T$-seconds of local $t$-time 
at some spatial location will take exactly $T$-seconds of local $t$-time, if it would happen at another location.
The fact that we can write the covariant Maxwell's equations in the simple form (\ref{eq:MaxwellReduced}) 
and not worry about the choice of the spatial coordinate system is 
due to the spatial flatness of the FRW metric, which we assumed at the beginning. 
Any (flat, stationary, no stretching, etc.) rectangular Cartesian system would do \cite{LandauLifshitz}.

Now, we introduce a variable change
\begin{align}
\label{eq:TauTime}
\tau(t)=\int\limits_{t_{0}}^{t}\frac{{\rm d}t'}{a(t')},
\end{align}
where $t_{0}$ is the ``switch-on'' moment of the causal source $\bJ(\bx,t)$.
The new time-like variable is known as the {\it cosmological time}, and the corresponding 
variable change procedure is sometimes called a {\it conformal transformation} \cite{Debate}, 
since it transforms the general-relativistic FRW metric into the Minkowski 
metric of special relativity. Indeed, in terms of this new variable the Maxwell equations become
\begin{align}
\label{eq:MaxwellTau}
\begin{split}
-\nabla\times\bB(\bx,\tau)+\mu_{0}\partial_{\tau}\bD(\bx,\tau)&=-\mu_{0}a^{4}(\tau)\bJ(\bx,\tau),
\\
\nabla\times\bD(\bx,\tau)+\varepsilon_{0}\partial_{\tau}\bB(\bx,\tau)&=0,
\end{split}
\end{align}
and are the same as Maxwell's equations in Minkowski vacuum up to the right-hand side and 
the above variable change. At this moment we still cannot say whether
the source is receeding with respect to the observation point or not, as the Maxwell
equations represent a {\it local} relation between the current density and the fields
at the same point in space. To make any conclusions about the possible
physical recession we need an explicit solution of these equations relating
the current density at one location to the fields at some other location.

From the mathematical point of view, the right-hand side of (\ref{eq:MaxwellReduced}) 
contains functions of $\tau$ which must be interpreted as $a(t(\tau))$ and $\bJ(\bx,t(\tau))$, 
where $t(\tau)$ is a function inverse with respect to $\tau(t)$. Obviously, this inverse function 
is well-defined whenever $a(t)>0$ for all $t\ge t_{0}$. 
Equations (\ref{eq:MaxwellTau}) are identical in form to the vacuum Maxwell's 
equations and therefore have a well-known analytical solution. In particular, they can be reduced 
\cite{Budko2009b} to the following scalar wave equation:
\begin{align}
\label{eq:ScalarWave}
\begin{split}
&\Delta\bD(\bx,\tau)-\frac{1}{c_{0}^{2}}\partial_{\tau}^{2}\bD(\bx,\tau)
=
\\
&\frac{1}{c_{0}^{2}}\partial_{\tau}\left[a^{4}(\tau)\bJ(\bx,\tau)\right]
-\nabla(\nabla\cdot)\int\limits_{0}^{\tau}a^{4}(\tau')\bJ(\bx,\tau')\,{\rm d}\tau'.
\end{split}
\end{align}
Now we can see the crucial difference between the present approach and that of Schr{\"o}dinger \cite{Schrodinger1939}.
Although the left-hand side of (\ref{eq:ScalarWave}) is formally the same as the one obtained 
in \cite{Schrodinger1939}, Schr{\"o}dinger had to use a different variable change to arrive at it.
Indeed, the covariant approach to a scalar wave equation 
would have resulted in $a^{3}(t)$ in the denominator of (\ref{eq:TauTime}), whereas
what we have used here is in perfect agreement with the usual definition of the 
cosmological time, represents a conformal transformation, and appears in 
other first-order formulations of the problem \cite{Plebanski1960, deFelice1971, Bahram1973}.
The explicit solution of (\ref{eq:ScalarWave}) in $(\bx,\tau)$-domain is:
 \begin{align}
 \label{eq:XTauEField}
 \begin{split}
 \bE(\bx,\tau)=
 &\int\limits_{\bx'\in{\mathbb R}^{3}}\frac{\left[3{\mathbb Q}-{\mathbb I}\right]}{4\pi\vert\bx-\bx'\vert^{3}}
 \int\limits_{0}^{\tau_{\rm r}}\frac{a^{4}(\tau')\bJ(\bx',\tau')}{\varepsilon_{0}a(\tau)}\,{\rm d}\tau'\,{\rm d}\bx'
 \\
 +&\int\limits_{\bx'\in{\mathbb R}^{3}}\frac{\left[3{\mathbb Q}-{\mathbb I}\right]}{4\pi\vert\bx-\bx'\vert^{2}}
 \frac{a^{4}(\tau_{\rm r})\bJ(\bx',\tau_{\rm r})}{c_{0}\varepsilon_{0}a(\tau)}\,{\rm d}\bx'
 \\
 +&\int\limits_{\bx'\in{\mathbb R}^{3}}\frac{\left[{\mathbb Q}-{\mathbb I}\right]}{4\pi\vert\bx-\bx'\vert}
 \frac{\left.\partial_{\tau}\left[a^{4}(\tau)\bJ(\bx',\tau)\right]\right\vert_{\tau=\tau_{\rm r}}}
 {c_{0}^{2}\varepsilon_{0}a(\tau)}\,{\rm d}\bx',
 \end{split}
 \end{align}
In this formula 
\begin{align}
\label{eq:Tensors}
\left[{\mathbb Q}(\bx-\bx')\right]_{ij}=\frac{(x_{i}-x'_{i})(x_{j}-x'_{j})}{\vert\bx-\bx'\vert^{2}},
\end{align}
and ${\mathbb I}$ denotes the  Kronecker $(3\times 3)$-identity tensor.
The retarded $\tau$-time is
 \begin{align}
 \label{eq:RetardedTau}
 \tau_{\rm r}=\tau-\frac{\vert\bx-\bx'\vert}{c_{0}}.
 \end{align}
Apart from the source modulation (discussed below), the obtained $(\bx,\tau)$-domain
solution is mathematically identical to the $(\bx,t)$-domain free-space radiation formula \cite{Budko2009b}. 
It does not provide any clear evidence of the mutual recession between the source at $\bx'$ and the observer at $\bx$.
In fact, there are only two possibilities for the recession to enter our solution: either in the form of the
co-moving distance $a(\tau)\vert\bx-\bx'\vert$ in terms of the comsological time $\tau$,
or in terms of the synchronous time as $a(t)\vert\bx-\bx'\vert$.
If the recession is really happenning, then this co-moving distance should appear everywhere,
where one would normally have a stationary distance $\vert\bx-\bx'\vert$, i.e. both in the distance-dependent 
decay factors and in the retarded time expression. At the moment we can already dismiss the first of 
the above possibilities. Of course, we could artificially group the $a(\tau)$ terms in the denominator with
the corresponding $\vert\bx-\bx'\vert$ terms. However, as shown below, this is mainly influencing the magnitude 
of the received field. For the recession to really {\it cause} the redshift the co-moving distance should be present in 
the temporal argument of the current density, i.e. in the retarded $\tau$-time (\ref{eq:RetardedTau}), 
which is obviously not the case. 

If there is no mutual recession, where does the redshift come from?
Mathematically, the reason for the cosmological redshift may be formulated as follows. The (electromagnetic) 
processes that are synchronous in the local synchronous time $t$ are not synchronous in the cosmological 
time $\tau$, and vice versa. To show this explicitly we need to re-write the $(\bx,\tau)$-domain radiation formula
(\ref{eq:XTauEField}) as an $(\bx,t)$-domain formula, which requires an inverse transformation $t(\tau)$.
For a general expansion rate $a(t)>0$ this inverse transformation exists but is implicit.
For the linear case, $a(t)=a_{0}+bt$, however, the integral (\ref{eq:TauTime}) can be evaluated
and the resulting algebraic equation can be solved for $t$ as:
\begin{align}
\label{eq:TasTau}
t=\frac{a_{0}+bt_{0}}{b}e^{b\tau}-\frac{a_{0}}{b},
\end{align}
which also means 
\begin{align}
\label{eq:ATau}
a(\tau)=\left(a_{0}+bt_{0}\right)e^{b\tau}.
\end{align}
The most obvious demonstration of the redshift comes from considering a monochromatic
source \cite{Budko2009b}. In that case the right-hand 
side of (\ref{eq:XTauEField}) contains trigonometric functions, say $\sin(\omega t)$. 
In view of (\ref{eq:TasTau}) and (\ref{eq:RetardedTau}) this functions will enter (\ref{eq:XTauEField}) as
\begin{align}
\label{eq:RetardedFunction}
\begin{split}
\sin(\omega t(\tau_{\rm r}))
&=\sin\left(\omega\left(\frac{a_{0}+bt_{0}}{b}e^{b(\tau-\vert\bx-\bx'\vert/c_{0})}-\frac{a_{0}}{b}\right)\right),
\end{split}
\end{align}
or, recognizing here the $t$-domain result, as
\begin{align}
\label{eq:RetardedFunction1}
\begin{split}
\sin\left(\omega' t+\frac{a_{0}}{b}(\omega'-\omega)\right),
\end{split}
\end{align}
where the new frequency is 
\begin{align}
\label{eq:PrimedTime}
\omega'=\omega e^{-\frac{b}{c_{0}}\vert\bx-\bx'\vert}.
\end{align}
Hence, a monochromatic signal sent at frequency $\omega$ from $\bx'$ will be observed at $\bx$
at a lower frequency $\omega'$. Moreover, this cosmological redshift will have an exact
exponential dependence on the distance between the source and the observer. 
At this point we can dismiss the other possibility for the mutual recession between the source
and the observer. We have just analyzed the retarded time expression in the $(\bx,t)$-domain
and dot see any signs of the co-moving distance $a(t)\vert\bx-\bx'\vert$ in it.
On the bright side, the present interpretation is completely free from any conceptual problems
as far as redshifts larger than one are concerned. Since their origin is clearly not kinematic, 
no superluminal velocities are required to explain them. It is also interesting to note 
that the redshift is completely independent of the inital moment of expansion, as long as this
initial moment was before or exactly at $t_{0}$ -- the moment at which the source starts radiating.

This redshift effect is a case of the general {\it time dilation}, which is confirmed by the observed
apparent change in the duration of various well-understood astronomical processes. 
Let the two characteristic time instants in the evolution of the source current density be $t_{1}^{\rm src}$ and 
$t_{2}^{\rm src}$. The local-time interval at the source location $\bx'$ is then
\begin{align}
\label{eq:DeltaTsrc}
\Delta t^{\rm src}= t_{1}^{\rm src}-t_{2}^{\rm src}=
\frac{a_{0}+bt_{0}}{b}\left[e^{b\tau_{1}^{\rm src}}-e^{b\tau_{2}^{\rm src}}\right],
\end{align}
where $\tau_{1}^{\rm src}$ and $\tau_{2}^{\rm src}$ are the corresponding $\tau$-times.
To derive the corresponding local times at the observer location 
we apply (\ref{eq:RetardedTau}) in (\ref{eq:TasTau})
and obtain:
\begin{align}
\label{eq:DeltaTsrc}
\Delta t^{\rm rec}&= t_{1}^{\rm rec}-t_{2}^{\rm rec}=
\frac{a_{0}+bt_{0}}{b}
\left[e^{b\tau_{1}^{\rm src}} - e^{b\tau_{2}^{\rm src}}\right]e^{\frac{b}{c_{0}}\vert\bx-\bx'\vert}.
\end{align}
Hence, it follows that
\begin{align}
\label{eq:Redshift}
\frac{\Delta t^{\rm rec}}{\Delta t^{\rm src}}=e^{\frac{b}{c_{0}}\vert\bx-\bx'\vert}.
\end{align}
Thus, the general time-dilation in a flat universe linearly expanding at a constant 
rate $b$ is an exponential function of the distance between the source and the observer.
In \cite{Budko2009b} the redshift formula for an exponentially expanding (de Sitter) universe is 
derived. Similarly to the linear case, it is a function of distance which mathematically does 
not even resemble the expansion rate function $a(t)$.

So far we have considered just one effect of the time-variation of the 
metric -- the cosmological redshift. In addition to that the obtained analytical solution 
(\ref{eq:XTauEField}) shows the modulation of the source by the fourth power of the expansion rate. 
This demonstrates that there is more to the time-varying metric than a simple change of the rate 
of emission due to the time dilation.
Substitution of (\ref{eq:ATau}) into (\ref{eq:XTauEField}) with the retarded
time (\ref{eq:RetardedTau}) will produce a multiplier of the form
\begin{align}
\label{eq:RetardedATau}
a(\tau_{\rm r})=\left(a_{0}+bt_{0}\right)e^{b\tau_{\rm r}}=a(\tau)e^{-\frac{b}{c_{0}}\vert\bx-\bx'\vert}.
\end{align}
Since this term comes in the fourth power, the additional decay in the luminocity is by 
far stronger than would follow from the reduced emission rate. 
In fact, the sources are disappearing from our view
significantly faster than they get redshifted.
This means that the exact redshift-luminocity relation for a linearly 
expanding universe, which still needs to be wroked out, is probably also different 
from the currently used one, especially at higher redshifts. 

The modulation of the current produces another curious effect. 
We know that a DC-like current does not normally radiate.
This follows from the fact that the far-field term for a vacuum background 
contains the time-derivative of the current. However, now the current is 
multiplied by the expansion rate. Thus, a DC-like current in a time-varying metric
will radiate into the far-field zone. On the other hand, a time-varying current,
which normally does radiate, will be effectively silenced, if its time variation
happens to coinside with $a^{-4}(\tau)$. 

In conclusion, the derived analytical radiation formula for an arbitrary source in an
infinite, flat, linearly expanding universe provides the exact exponential redshift/time-dilation 
law valid for all distances and free from intepretational problems for higher redshifts. 
It appears that both the dime-dilation and the decrease in the apparent brightness of the sources 
caused by the interaction of light with a time-varying FRW metric have 
so far been greatly underestimated. The obtained solution provides no convincing evidence
for the change of the distances between the source and the observer in the FRW cosmology. 
Paradoxically, expanding universe does not follow from the expanding metric.

%%%%%%%%%%%%%%%%%%%%%%%%%%%%%%%%%%%%%%%%%%%%%%%%%%%%%%%%%%%%%%%%%%%%%%%%%%%%
%%

\end{document}